# Ultra-Reliable Low Latency Cellular Networks: Use Cases, Challenges and Approaches

He Chen, Rana Abbas[1], Peng Cheng[1], Mahyar Shirvanimoghaddam[1], Wibowo Hardjawana[1], Wei Bao[1], Yonghui Li, and Branka Vucetic

The University of Sydney, NSW 2006, Australia
Email: firstname.lastname@sydney.edu.au

*Abstract*-The fifth-generation cellular mobile networks are expected to support mission critical ultra-reliable low latency communication (URLLC) services in addition to the enhanced mobile broadband applications. This article first introduces three emerging mission critical applications of URLLC and identifies their requirements on end-to-end latency and reliability. We then investigate the various sources of end-to-end delay of current wireless networks by taking the 4G Long Term Evolution (LTE) as an example. Subsequently, we propose and evaluate several techniques to reduce the end-to-end latency from the perspectives of error control coding, signal processing, and radio resource management. We also briefly discuss other network design approaches with the potential for further latency reduction.

## I. Introduction

The growth of wireless data traffic over the past three decades has been relentless. The upcoming fifth-generation (5G) of wireless cellular networks is expected to carry 1000 times more traffic [1] while maintaining high reliability. Another critical requirement of 5G is ultra-low latency – the time required for transmitting a message through the network. The current fourth-generation (4G) wireless cellular networks have a nominal latency of about 50ms; however, this is currently unpredictable and can go up to several seconds [2]. Moreover, it is mainly optimized for mobile broadband traffic with target block error rate (BLER) of $10^{-1}$ before re-transmission.

---

[1] These authors contributed equally to this article.

There is a general consensus that the future of many industrial control, traffic safety, medical, and internet services depends on wireless connectivity with guaranteed consistent latencies of 1ms or less and exceedingly stringent reliability of BLERs as low as $10^{-9}$ [3]. While the projected enormous capacity growth is achievable through conventional methods of moving to higher parts of the radio spectrum and network densifications, significant reductions in latency, while guaranteeing an ultra-high reliability, will involve a departure from the underlying theoretical principles of wireless communications.

## II. Emerging URLLC Applications

In this section, we briefly introduce three emerging mission-critical applications, including tele-surgery, intelligent transportation, and industry automation, whose latency and reliability requirements will be identified. Other possible applications of URLLC include Tactile Internet, augmented/virtual reality, fault detection, frequency and voltage control in smart grids, which are not elaborated here due to space limitation.

*A. Tele-surgery*

The application of URLLC in tele-surgery has two main use cases [4]: (1) remote surgical consultations, and (2) remote surgery. The remote surgical consultations can occur during complex life-saving procedures after serious accidents with patients having health emergency that cannot wait to be transported to a hospital. In such cases, first-responders at an accident venue may need to connect to surgeons in hospital to get advice and guidance to conduct complex medical operations. On the other hand, in a remote surgery scenario, the entire treatment procedure of patients is executed by a surgeon at a remote site, where hands are replaced by robotic arms. In these two use cases, the communication networks should be able to support the timely and reliable delivery of audio and video streaming. Moreover, the haptic feedback enabled by various sensors located on the surgical equipment is also needed in remote surgery such that the surgeons can feel what the robotic arms are touching for precise decision-making. Among these three types of traffic, it is haptic feedback that requires the tightest delay requirement with the end-to-end round trip times (RTTs) lower than 1ms [4]. In terms of reliability, rare failures can be tolerated in remote surgical consultations, while the remote surgery demands an extremely reliable system (BLER down to $10^{-9}$) since any noticeable error can lead to catastrophic outcomes.

*B. Intelligent Transportation*

The realization of URLLC can empower several technological transformations in transportation industry [5], including automated driving, road safety and traffic efficiency services, etc. These transformations will get cars fully connected such that they can react to increasingly complex road situations by cooperating with others rather than relying on their local information. These trends will require information to be disseminated among vehicles reliably within extremely short time duration. For example, in fully automated driving with no human intervention, vehicles can benefit by the information received from roadside infrastructure or other vehicles. The typical use cases of this application are automated overtake, cooperative collision avoidance and high density platooning, which require an end-to-end latency of 5–10ms and a BLER down to $10^{-5}$ [5].

*C. Industry Automation*

URLLC is one of the enabling technologies in the fourth industrial revolution [6]. In this new industrial vision, industry control is automated by deploying networks in factories. Typical industrial automation use cases requiring URLLC include factory, process, and power system automation. To enable these applications, an end-to-end latency lower than 0.5ms and an exceedingly high reliability with BLER of $10^{-9}$ should be supported [3]. Traditionally, industrial control systems are mostly based on wired networks because the existing wireless technologies cannot meet the industrial latency and reliability requirements. Nevertheless, replacing the currently used wires with radio links can bring substantial benefits: (1) reduced cost of manufacturing, installation and maintenance; (2) higher long-term reliability as wired connections suffer from wear and tear in motion applications; (3) inherent deployment flexibility.

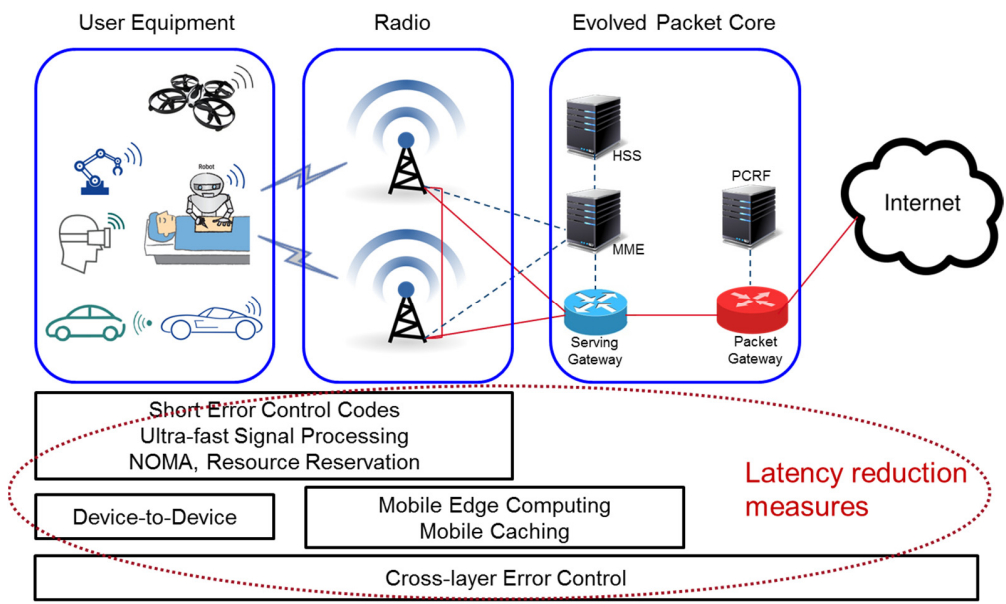

Fig. 1. Architecture of 4G LTE network with representative mission-critical user equipment. The bottom part lists various potential measures towards latency reduction in different parts.

## III. Latency Sources in Cellular Networks

Cellular networks are complex systems with multiple layers and protocols, as depicted in Fig. 1. The duration of a data block at the physical layer is a basic delay unit which gets multiplied over higher layers and thus causes a considerable latency in a single link. On the other hand, protocols at higher layers and their interactions are significant sources of delay in the whole network.

Latency varies significantly as a function of multiple parameters, including the transmitter–receiver distance, wireless technology, mobility, network architecture, and the number of active network users.

TABLE I

VARIOUS DELAY SOURCES OF AN LTE SYSTEM (RELEASE 8) IN THE UPLINK AND DOWNLINK

| Delay Component | Description | Time (ms) |
|---|---|---|
| Grant acquisition | A user connected and aligned to a base station will send a Scheduling Request (SR) when it has data to transmit. The SR can only be sent in an SR-valid Physical Uplink Control Channel (PRCCH). This component characterizes the average waiting time for a PRCCH. | 5ms |
| Random Access | This procedure applies to the users not aligned with the base station. To establish a link, the user initiates an uplink grant acquisition process over the random access channel. This process includes preamble transmissions and detection, scheduling, and processing at both the user and the base station. | 9.5ms |
| Transmit time interval | The minimum time to transmit each packet of request, grant or data | 1ms |
| Signal processing | The time used for the processing (e.g., encoding and decoding) data and control | 3ms |
| Packet retransmission in access network | The (uplink) hybrid automatic repeat request process delay for each retransmission | 8ms |
| Core network/Internet | Queueing delay due to congestion, propagation delay, packet retransmission delay caused by upper layer (e.g., TCP) | Vary widely |

The latency components of the LTE networks have been systematically evaluated and quantified by 3rd Generation Partnership Project (3GPP) [7]. Latencies for various radio access network algorithms and protocols in data transmission from a user to the gateway (i.e., uplink) and back (i.e., downlink) are summarized in Table I. The two most critical sources of delay in radio access networks are the link establishment (i.e., grant acquisition or random access) and packet retransmissions caused by channel errors and congestion. Another elementary delay component is the transmit time interval (TTI), defined as the minimum data block length, which is involved in each transmission of grant, data, and retransmission due to errors detected in higher layer protocols.

According to Table I, after a user is aligned with the base station, its total average radio access delay for an uplink transmission can be up to 17ms excluding any retransmission, which includes the following steps: UE waits for a Physical Uplink Control Channel (5ms) → UE sends a scheduling request (1ms) → BS decodes the scheduling the scheduling request and generates the scheduling grant (3ms) → BS sends the scheduling grant (1ms) → UE decodes the scheduling grant (3ms) → UE sends uplink data (1ms) → BS decodes the data (3ms). On the other hand, each downlink data transmission will include the following procedures: incoming date processing (3ms), TTI alignment (0.5ms), transmission of the downlink data (1ms), data decoding in UE (3ms), which sums up to 7.5ms and is lower than that of the uplink since no grant acquisition process is needed in the downlink. The overall end-to-end latency in cellular networks is dictated not only by the radio access network but also includes delays of the core network, data center/cloud, Internet server and radio propagation. It increases with the transmitter-receiver distance and the network load. As shown by the experiment conducted in [8], at least 39ms is needed to contact the core network gateway, which connects the LTE system to the Internet, while a minimum of 44ms is required to get response from the Google server. As the number of users in the network rises, the delay goes up, due to more frequent collisions in grant acquisition and retransmissions caused by inter-user interference.

In the subsequent sections, we will consider novel approaches that could be implemented at various cellular network layers (as depicted in the bottom part of Fig. 1) to support ultra-low latency services. We note that the sources of delay listed in Table I can be grouped into five categories: (1) TTI, (2) signal processing, (3) Radio resource management, (4) Retransmissions, and (5) Core network. In the remainder of the article, we will focus mainly on how to reduce the first three delay categories in Sec. IV, V and VI, respectively, while we will briefly discuss some ideas that can impact on the last two categories in Sec. VII. It is worth mentioning that a very recent work [9] also proposed several promising solutions to achieve low latency for enabling URLLC, from the perspective of waveform design, radio slot structure, radio resource management and channel access, etc.

## IV. Short Error Control Codes

In traditional communication systems, very long low-density parity check (LDPC) or turbo codes are used to achieve near error-free transmissions, as long as the data rate is below the Shannon channel capacity. Since the network latency is significantly affected by the size of data blocks, short codes, corresponding to shorter TTI, are a prerequisite for low delays; but the Shannon theoretical model breaks down for short codes. A recent Polyansky-Poor-Verdu (PPV) analysis of channel capacity with finite block lengths [10] has provided the tradeoffs between delays, throughput, and reliability on Gaussian channels and fixed rate block codes, by introducing a new fundamental parameter called 'channel dispersion'; this analysis shows that there is a severe capacity loss at short block-lengths. There are no known codes that achieve the PPV limit. Low-density parity check (LDPC) codes and polar codes have been reported to achieve almost 95% of the PPV bound at block error rates as low as $10^{-7}$ for block lengths of a few hundred symbols [11]. However, their main drawback is the large decoding latency.

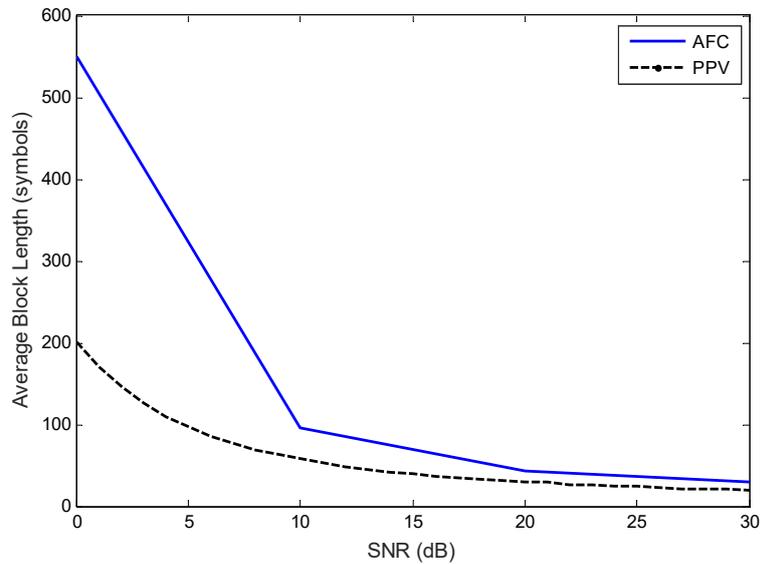

Fig. 2. Analog fountain code (AFC) with a 0.95-rate Protograph-based LDPC precoder are used to encode a message of length 192 bits for a block error rate of $10^{-4}$ over a wide range of SNRs for the AWGN channel. PPV represents Polyansky-Poor-Verdu.

As long fixed rate codes achieve the Shannon capacity limit for one signal-to-noise ratio (SNR) only, today's wireless networks use adaptive schemes, which select a code from a large number of fixed rate codes, to transmit data at the highest possible rate for a specified reliability and estimated channel state information (CSI). The problem is the inevitable latency increase due to complex encoding and decoding algorithms, the time required to estimate the CSI at the receiver, the feedback of CSI back to the transmitter, code rate and modulation selection process in the transmitter, and block length.

In this context, self-adaptive codes appear as a promising solution to URLLC. Self-adaptive codes, also known as rateless codes, can adapt the code rate to the channel variations by sending an exact amount of coded symbols needed for successful decoding. This self-adaptation does not require any CSI at the transmitter side, thus eliminating the channel estimation overhead and delay. Recently, an analog fountain code (AFC) [12] was proposed as a capacity-approaching rateless code over a wide range of SNRs for asymptotically long codewords. AFC can be represented by a single sparse non-binary generator matrix such that the optimization of the coding and modulation can be performed jointly via specialized EXIT charts. The resulting performance is seamless over a large range of SNRs with only linear encoding and decoding complexity with respect to the block length. In Fig. 2, we show that AFC, even in the current sub-optimal design for short codes, has a small gap to the PPV bound in the medium to high SNR regime. Moreover, we expect that a much lower latency can be achieved when optimizing AFC for shorter block lengths. As self-adaptive codes do not require any CSI to be available at the transmitter side, the channel estimation overhead can be eliminated.

Note that the implementation of AFC will require the receiver to send a positive feedback once a codeword has been successfully decoded. Such a positive feedback can be sent in an acknowledgement packet to be transmitted in UL (or DL) slots. Furthermore, the feedback can suffer from packet loss in practice. According to the principle of AFC, if the positive feedback is not received, the transmitter will continue to transmit the same codeword, which will introduce extra delay. As such, more robust coding scheme should be applied on the acknowledgment packet to avoid such extra delay.

## V. Ultra-fast Signal Processing

The current LTE systems use system throughput as the main design target and performance indicator. In contrast, signal processing latency issues has drawn far less attention in the design process. Similar to Section III, valuable insights into the processing latency bottleneck in the current LTE systems could be obtained by a breakdown of latencies contributed by each LTE receiver module. To this end, we investigate the average computational time for the major receiver modules of an LTE Release 8 system by implementing it on an Intel Core i5 computer.

The computational time, a practical indicator for relative latency, is presented in Table II for three typical bandwidths. In the simulations, we have 4 transmit and 2 receive antennas, 16-QAM, and 0.3691 code rate at signal-to-noise ratio of 10dB. The closed-loop spatial multiplexing mode was implemented and the average computational time is based on one subframe. It is clearly shown that MMSE-based channel estimation, MMSE-SIC-based MIMO detection, and Turbo decoding consume the most computational resources and dominate the computational time. To lower the processing latency, new ultra-fast signal processing techniques, especially for the three identified functions, should be developed to strike a favorable tradeoff between throughput and latency.

TABLE II
A COMPARISON OF COMPUTATIONAL TIME FOR DIFFERENT FUNCTION MODULES AT THE RECEIVER, WHEREIN ALL NUMBERS WITHOUT A UNIT ARE IN SECONDS. THROUGHPUT IS DERIVED FROM THE OBTAINED BLER IN THE LTE SIMULATION PLATFORM.

| Receive Modules | B = 1.4MHz | B = 5MHz | B = 10MHz |
|---|---|---|---|
| CFO Compensation | 0.0010 | 0.0023 | 0.0037 |
| FFT | 2.9004e-04 | 6.2917e-04 | 8.3004e-04 |
| Disassemble Reference Signal | 1.2523e-04 | 2.2708e-04 | 3.1685e-04 |
| Channel Estimation (MMSE) | 0.0015 | 0.0141 | 0.0878 |
| Disassemble Symbols | 0.0013 | 0.0045 | 0.0087 |
| MIMO Detection (MMSE-SIC) | 0.0028 | 0.0242 | 0.0760 |
| SINR Calculation | 2.4947e-04 | 6.6754e-04 | 0.0012 |
| Layer Demapping | 4.3253e-05 | 1.0988e-04 | 3.8987e-04 |
| Turbo Decoding | 0.0129 | 0.0498 | 0.1048 |
| Obtained Throughput | 2.2739Mbps | 10.073Mbps | 20.41Mbps |

In our simulation, we propose and implement an improved channel estimation approach to reduce the channel estimation latency. The basic idea is to use the least square estimation to extract the CSI associated with the reference symbols, and then employ an advanced low-complexity 2-D biharmonic interpolation method to obtain the CSI for the entire resource block. Typically, the resulting curves from the biharmonic interpolation method are much smoother than the linear and nearest neighbor methods. Our simulation results show that the proposed channel estimation method can reduce around 60% of the computational time relative to the MMSE-based method at B = 5MHz, while achieving almost the same system throughput.

It is also desirable to develop ultra-fast multilayer interference suppression technologies to enable fast MIMO detection, especially for a large number of transmit and receive antennas. Along this direction, a parallel interference cancellation (PIC) with decision statistical combining (DSC) detection algorithm was developed in [13], which can significantly reduce the detection latency compared with MMSE-SIC. The PIC detectors are equivalent to a bank of matched filters, which avoid the time-consuming MMSE matrix inversion. A very small number of iterations between the decoder and the matched filter are added to achieve the performance of MMSE receivers. This algorithm was also applied to ICI cancellation for high-mobility MIMO-OFDM systems. In a typical configuration, PIC-DSC can reduce the computational complexity by 128 times compared to MMSE and ZF, with a negligible performance degradation.

Parallel hardware implementation is another important measure to reduce signal processing latency. For example, the recently proposed parallel turbo decoder architecture [14] eliminates the serial data dependencies, realizes full parallel processing, realizes full parallel processing, offers an average throughput of 1.53 Gb/s, and finally achieves a 50% hardware resource reduction compared with the original architecture.

# VI. Radio Resource Management

In this section, we will discuss two radio resource management techniques that have great potential to reduce the latency caused by the medium access process.

*A. Non-orthogonal Multiple Access*

As shown in Table I, grant acquisition and random access procedures in current standards are two major sources of delay. This calls for novel approaches and fundamental shifts from current protocols and standards originally designed for human communication to meet the requirements for ultra-low latency applications. Though optimal in terms of per user achievable rate, orthogonal multiple access (OMA) techniques, such as OFDMA in current LTE, are major causes of the latency associated with the link establishment and random access. More specifically, in existing wireless systems, radio resources are orthogonally allocated to the users to deliver their messages. This requires the base station to first identify the users through contention-based random access. This strategy suffers from severe collisions and high latencies when the number of users increases.

Non-orthogonal multiple access (NOMA) has recently gained considerable attention as an effective alternative to conventional OMA. Generally, NOMA can be further categorized into two types: power-domain NOMA and code-domain NOMA. In this article, we consider the code-domain NOMA. Specifically, each user is assigned with a unique channel code, which is used to identify different users instead of the power levels used in power-domain NOMA. In this case, the users in code-domain NOMA are enabled to transmit in a grant-free manner. Furthermore, the adopted channel codes can also boost the reliability of the users [15].

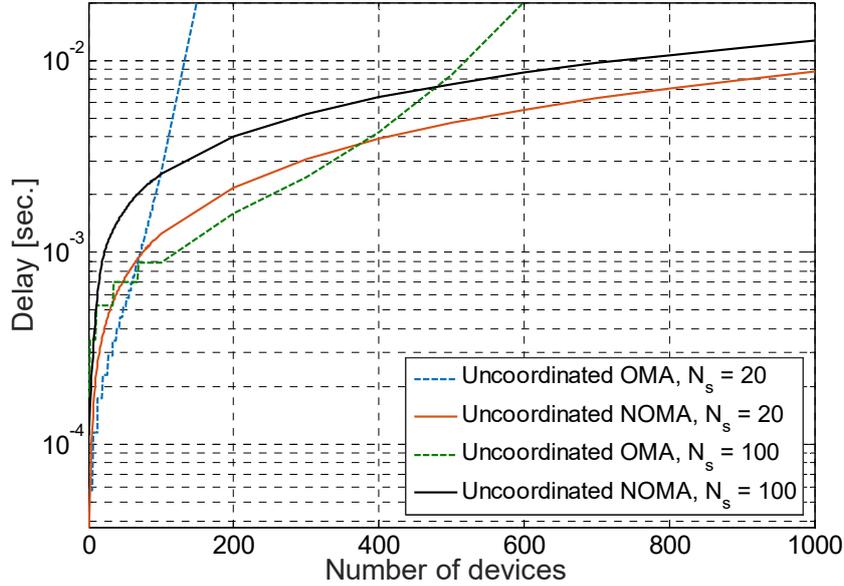

Fig. 3. Delay versus the number of devices for NOMA and OMA.

Fig. 3 shows a comparison between NOMA and OMA in an uncoordinated scenario, where the devices randomly choose a subband for their transmission. The number of subbands is denoted by $N_s$ and the total available bandwidth is assumed to be W = 100MHz. The bandwidth is assumed to be uniformly divided into $N_s$ subbands, each of W/$N_s$ bandwidth. As can be seen, when the number of devices is small, OMA slightly outperforms NOMA in terms of delay, which is expected as the collision probability in this case is small and the devices can achieve higher spectral efficiency as they are transmitting orthogonally. However, when the number of devices is large, NOMA outperforms OMA, as it can effectively exploit the interference and enable the devices to be decoded at the base station. In other words, in high traffic load scenarios, OMA is mainly dominated by the random access collision which leads to unavoidable high latencies, while NOMA supports a large number of devices with the desired latency, by eliminating the random access phase and enabling the users to share the same radio resources.

The main benefits of NOMA come from the fact that it does not need separate grant acquisition and random access phase, as the devices can send their data whenever they want to send. This becomes more beneficial when the number of devices grows large, which is the scenario of interest for most Internet-of-Things use cases. NOMA can also be easily combined with AFC codes [12] to improve the spectral efficiency and reliability for each user, therefore providing a cross-layer solution for reducing the delay. Besides, it is worth emphasizing that the choice of the channel access method strongly depends on the application scenarios. For the URLLC scenarios with the relatively fixed users (e.g., factory automation), contention-based access is not necessary and pre-allocation-based access is more efficient. On the other hand, for those URLLC scenarios with a possibly large number of users joining and leaving the network frequently (e.g., intelligent transportation systems), contention-based access is essential and thus NOMA could be more beneficial.

*B. Resource Reservation via Resource Block Slicing*

In the current LTE network, the management of radio resource blocks (RBs) for multiple services is jointly optimized. As such, the latencies of different services are interdependent. A traffic overload generated by one service can negatively impact the latency performance of other services. To address this issue, we propose to reserve radio resources for each service. The reservation is done by slicing RBs and allocating a slice to each service based on the traffic demand. Moreover, if RBs in a slice are not used, they will be shared by other services. This type of resource reservation method can achieve a high spectral efficiency and eliminate the latency problem caused by the traffic overload issues coming from other services.

To evaluate the benefit of the proposed RB slicing on a LTE network, we conduct a simulation to compare its performance with a legacy LTE network by using NS-3. Two types of services with different data rates and latency requirements, i.e., low latency intelligent transportation systems (ITS) with average packet sizes of 100 bytes and average packet intervals of 100ms per user, and smart grid (SG) with average packet sizes of 300 bytes and average packet intervals of 80ms per user, respectively, are considered in our simulation. The devices for the above services are distributed in 1 km$^2$ area according to a Poisson Point Process (PPP) with averages of 400 and 600 devices for ITS and SG, respectively, served by 4 LTE base stations, operating with 20MHz bandwidth. Note that all available RBs are shared by ITS and SG equally in the current LTE network. In the simulation, the unused RBs sliced to ITS devices can be used by SG devices, and vice versa.

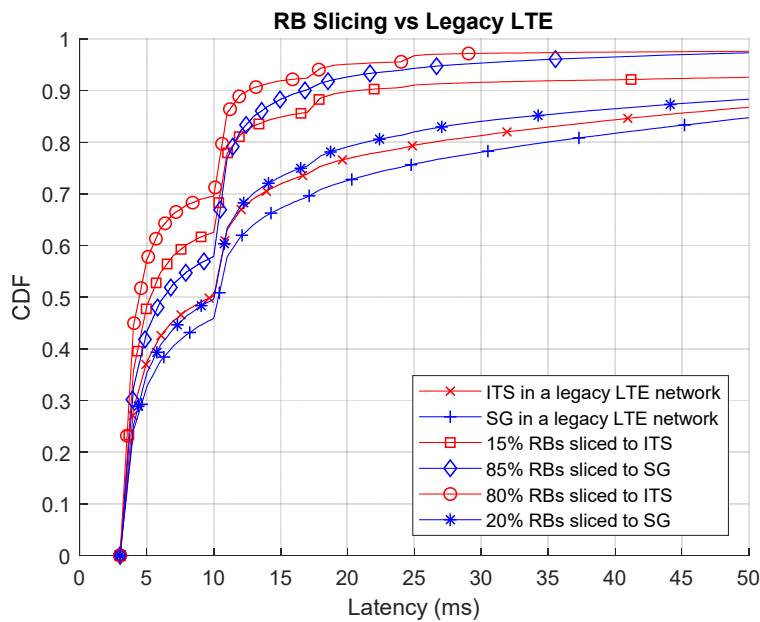

Fig. 4. The cumulative distribution function (CDF) of the end-to-end delay without and with radio resource block slicing. [The authors would like to thank Zhouyou Gu for his assistance in simulating this figure]

Fig. 4 shows the cumulative density function (CDF) for the end-to-end packet latencies under a legacy LTE network and under the RB slicing regime that isolates the traffic demand of ITS and SG from each other. We can see from Fig. 4 that when the proportion of reserved RBs for ITS and SG devices are set to match the traffic proportion of their devices to the total traffic generated by all devices, namely 15% and 85%, the latency performance of those two slices are almost the same. Note that the generated ITS traffic is 3.2 Mbps (400 devices sending 100 bytes every 100 ms), while the generated SG traffic is 18 Mbps (600 devices sending 300 bytes every 80 ms). More specifically, by performing RB slicing that reserves resources for each service, the latency is reduced from a median of 10ms to 5ms and 6ms for ITS and SG devices, respectively. We also observe if the resources reserved for ITS and SG devices are now set to different numbers, namely 80% and 20%, the latency performance of ITS devices becomes better. This is because more RBs are allocated to ITS devices. On the other hand, the latency of the SG devices deteriorates since the RBs sliced to them decreases from 85% to 20%, which is, however, still better than that in a legacy LTE network without RB slicing. This simulation confirms the benefit of the proposed resource reservation approach.

## VII. Other Potential Techniques

In addition to the measures introduced in previous sections, there are other techniques that have great potential to reduce the end-to-end latency of cellular systems. In what follows, we briefly discuss the principles of five potential technologies and explain how they can reduce latency.

**Cross-layer Error Control:** User Datagram Protocols (UDP), with no Automatic Repeat reQuest (ARQ) retransmissions and lower overheads than Transport Control Protocol (TCP), is attractive for emerging mission-critical communications over wireless networks. However, in order for UDP to be suitable for URLLC, its reliability needs to be significantly improved. A promising solution to resolve this is to use short AFC codes in both the physical and the network layer and form a concatenated code with soft output decoding at the physical and soft input decoding at the network layer. Furthermore, the decoding of both AFC codes can be highly parallelized for a low decoding delay.

**Device-to-Device Communication**: Device-to-device (D2D) communication refers to a radio technology that enables direct communication between two physically close terminals. D2D has recently been considered as a key solution for ultra-low latency applications, as it provides a direct link between traffic participants, without going through the network infrastructure. Due to the global spectrum shortage, D2D links are expected to operate within the same spectrum used by existing infrastructure-based communication systems (e.g., cellular systems). This calls for highly efficient interference management techniques to ensure the harmonious coexistence between D2D links and conventional links. Otherwise, the latency gain introduced by D2D communication can easily disappear.

**Mobile Edge Computing:** Mobile edge computing (MEC) is a promising approach to promptly process computationally intensive jobs offloaded from mobile devices, thus reducing the end to end latency. Edge computing modules can be installed at base stations which are closer to sensing devices than data servers/clouds. The implementation of edge computing technologies is not mature in cellular networks. The key barrier stems from the incompatibility of computing services and the existing LTE protocol stack. Modifying the existing stack to accommodate computing services may cause substantial network reconstruction and reconfiguration. Therefore, smoothly merging edge computing into the protocol stack is a key future research direction.

**Mobile Caching for Content Delivery**: Smart mobile caching schemes are also effective solutions for improving the delay performance of data intensive applications, e.g., multimedia, augment reality (AR) applications etc. Mobile caching enables content reuse, which leads to drastic delay reductions and backhaul efficiency improvements. The mobile cache can be installed at each base station. Whenever a mobile device's request "hits" a cached content, the base station intercepts the request and directly returns the cached content without resorting to a remote server. Despite the potential benefits of caching, it is still challenging to realize these benefits in practice. This is because the cache size at the base station is limited, but the number of possible contents can be unlimited. Thus, it is essential to determine how to wisely cache a set of popular contents to maximize the hit rate.

**Lightweight security mechanisms**: Security mechanisms are of great importance to URLLC applications since the transmitted information therein is normally critical and sensitive. Various cryptography techniques have been widely used in conventional communication systems to protect the transmitted information. However, cryptography techniques can considerably increase in message size due to the added overhead, leading to longer transmission latency. In this sense, it is essential to develop new lightweight security mechanisms (e.g., physical layer-based approaches) to guarantee the security of URLLC applications while introducing the minimal additional overhead (latency).

## VIII. Conclusions

This article has introduced the emerging applications, design challenges, and potential approaches in the design of ultra-reliable low latency communications (URLLC). We described potential use cases of URLLC in tele-surgery, smart transportation and industry automation and presented the latency and reliability requirements for these applications. To pinpoint major latency bottlenecks in current cellular networks, we showed a breakdown of the various delay sources in an LTE system and found that a few orders of end-to-end latency reduction is required to support the mission critical applications. To achieve this, each latency component needs to be reduced significantly. Our initial results showed that short analog fountain codes, ultra-fast signal processing, non-orthogonal multiple access and resource reservation via resource block slicing are essential to reduce latency in the physical and multiple access layers. Furthermore, other potential latency reduction measures, including cross-layer error control, device-to-device communication, mobile edge computing and mobile caching, were briefly discussed. We hope this article can encourage more research efforts toward the realization of URLLC.